\documentclass{llncs}

\usepackage{amsmath}
\usepackage{amsfonts}
\usepackage{amssymb}

\usepackage{enumerate}

\spnewtheorem*{theorem*}{Theorem}{\bfseries}{\rmfamily}
\spnewtheorem{claim}{Claim}{\itshape}{}

\newcommand{\SimS}{\mathsf{Sim}}
\newcommand{\Sim}[1]{\mathsf{Sim}\left(#1\right)}

\newcommand{\D}{\textsc{D}}
\newcommand{\A}{\textsc{A}}
\newcommand{\cD}{\mathcal{D}}
\newcommand{\cA}{\mathcal{A}}
\newcommand{\cC}{\mathcal{C}}
\newcommand{\F}{\textsc{F}}
\newcommand{\Var}{\mathrm{Var}}

\newcommand{\bigOm}[1]{\Omega\left(#1\right)}
\newcommand{\bigO}[1]{O\left(#1\right)}

\newcommand{\poly}{\mathrm{poly}}

\DeclareMathOperator*{\E}{\mathbb{ E }}

\usepackage{caption} 
\usepackage{tikz}
\usetikzlibrary{shapes,arrows,positioning, decorations.pathmorphing}

\usepackage{hyperref}
\usepackage[nameinlink]{cleveref}
\hypersetup{
  colorlinks=true,
  linkcolor=blue,
  citecolor=green
}
\crefname{table}{table}{tables}
\Crefname{table}{Table}{Tables}
\crefname{figure}{figure}{figures}
\Crefname{figure}{Figure}{Figures}
\crefname{section}{section}{sections}
\Crefname{section}{Section}{Sections}
\crefname{claim}{claim}{claims}
\Crefname{claim}{Claim}{Claims}

\title{A New Approximate Min-Max Theorem with Applications in Cryptography}

\author{Maciej Sk\'{o}rski 
\institute{
\email{maciej.skorski@mimuw.edu.pl} \\ Cryptology and Data Security Group, University of Warsaw}
}
\begin{document}

\maketitle

\begin{abstract}
We propose a novel proof technique that can be applied to attack a broad class of problems in computational complexity, when switching the order of universal and existential quantifiers is helpful. Our approach combines the standard min-max theorem and convex approximation techniques, offering quantitative improvements over the standard way of using min-max theorems as well as more concise and elegant proofs.
\end{abstract}

\section{Introduction}

\subsection{The Min-Max Theorem.}

The celebrate von Neumann min-max theorem~\cite{Neumann} states that every finite, two-player, zero-sum game has an equilibrium in mixed strategies. That is, the maximum value of the minimum expected gain for one player is equal to the minimum value of the maximum expected loss for the other. Any zero-sum game can be represented as a payoff matrix 
\begin{align*}
 A = \left[ A(x,y) \right]_{x\in X,y\in Y}
\end{align*}
where $A(x,y)$ is the payoff in case when the $X$-player chooses strategy $x \in X$ and the $Y$-player chooses strategy $y\in Y$, understood as a gain for the $X$-player and a loss of the $Y$-player. The basic moves $x\in X,y\in Y$ are called \emph{pure strategies} (think of one of 3 options in the rock-paper-scissors game). We allow the players to use randomized strategies, which are called \emph{mixed strategies} (think of picking a random answer in the rock-paper-scissors game) represented formally as distributions $p_X(\cdot),p_Y(\cdot)$ over $X$ and $Y$ respectively, and analyze the expected payoff 
\begin{align*}
\mathbb{E}_{y\sim p_Y,x\sim p_X} A(x,y) = \sum_{y}\sum_{x} p_Y(y)  p_X(x) A(x,y).
\end{align*}
If the player $X$ goes first, she can guarantee her gain to be at least
\begin{align*}
 \mathsf{MaxGain}(X) = \max_{p_X}\min_{y} \mathbb{E}_{x\sim p_X} A(x,y),
\end{align*}
and when the player $Y$ goes first he guarantees his lost to be at most 
\begin{align*}
 \mathsf{MinLoss}(Y) = \min_{p_Y}\max_{x} \mathbb{E}_{y\sim p_Y} A(x,y),
\end{align*}
where in both equations we used the fact that the second player always achieves the best response with some pure strategy. The min-max theorem guarantees that we have an equilibrium between the players.
\begin{theorem*}[Min-Max Theorem \cite{Neumann}]
With the notation as above (and players using mixed strategies), we have 
\begin{align*}
 \mathsf{MaxGain}(X) = \mathsf{MinLoss}(Y).
\end{align*}
\end{theorem*}
Many more general versions of the min-max theorem exist. All of them assure the equality
\begin{align*}
 \sup_{x\in X}\inf_{y\in Y} f(x,y) = \inf_{y\in Y}\sup_{x\in X} f(x,y)
\end{align*}
under certain conditions imposed on the sets $X,Y$ (for example both convex and compact subsets of a locally convex topological space)  and the function $f$ (for example continuity, convexity in $y$ and concavity in $x$). The proofs typically use fixed point theorems. Min-Max theorems have a lot of applications in game theory, statistical decision theory, economy and theoretical computer science. In this paper we focus on applications in cryptography, and the simplest version will be enough for our discussion.

\subsection{Switching the order of quantifiers by the min-max theorem} The min-max theorem may be used to change the order of quantifiers (minimization corresponds to the existential quantifier and maximization corresponds to the universal quantifier). A very good example is the classical hardcore lemma due to Impagliazzo \cite{Impagliazzo95hard-coredistributions}. The lemma stated informally says that if for every algorithm $A$ there exists a large set of inputs on which $A$ fails to compute a fixed function $f$, then in fact there exists a large set of inputs on which every algorithm fails to compute $f$ with probability close to $\frac{1}{2}$. This particular lemma falls into a broad class of results in complexity theory which can be proven using the min-max theorem. We explain this technique before giving more examples. 
\vspace{2mm}\newline\textsc{The general framework.}
Let $\cA$ be a class of test functions (for example poly-size circuits) over a set of possible inputs $I$ and $\cC$ be a class of distributions over $I$ satisfying certain desired properties (for example samplability, high density, high entropy etc.), and $v$ be a payoff function quantifies how well $A$ performs on the input $X$ (for example, unpredictability or distinguishing advantage). Suppose that we want to prove the existence of a distribution with certain properties for which every algorithm has bad (or alternatively good) performance.
\begin{quote}
\textbf{Dream Statement}. There is a distribution over inputs (with some certain properties) such that every algorithm performs badly/well. 
\begin{align}\label{eq:dream}
\exists X\in\mathcal{C} \ \forall \A\in \cA\: \quad v(\A,X) \leqslant c
\end{align}
\end{quote}
In many cases, it is much easier to prove a weaker version, which gives the existence of a distribution with desired properties but only for a chosen algorithm.
\begin{quote}
\textbf{Weak Statement}. For every algorithm there is a distribution over inputs (with some certain properties) such that it performs badly/well. 
\begin{align}\label{eq:weak}
\forall \A\in \cA\ \exists X\in\mathcal{C}: \quad v(\A,X) \leqslant c
\end{align}
\end{quote}
Note that this condition is considerably weaker. Indeed, we will see that in many applications proving the existence of a suitable distribution $X$ for a fixed algorithm $\A$ is actually trivial. But the big question is whether \Cref{eq:weak} implies \Cref{eq:dream} 
\begin{quote}
\textbf{Does the Weak Statement imply the Dream Statement?} Suppose that \Cref{eq:weak} holds. Can we conclude that \Cref{eq:dream} also holds, with possibly somewhat weaker class $\cA$ and a weaker parameter $c$?
\end{quote}
Note that we allow for some loss in quality (a weaker class of algorithms or a weaker payoff). Indeed, if both sets $\mathcal{C}$ and $\mathcal{A}$ are convex the answer is trivially ``yes'', by the min-max theorem. However, in most applications the set $\mathcal{A}$ consists of efficient algorithms (circuits of a bounded size) and is not convex, because taking a mixed strategy corresponds to combining many algorithms by (possibly) inefficient sampling. For the same reason, the set $\mathcal{C}$ might not be convex. However, we might ``embed'' non-convex sets $\mathcal{A}$ and $\mathcal{C}$ into ``almost'' convex hulls of $\cA',\cC'$ which are (hopefully) still sufficiently good for our purpose, by taking moderately long mixed strategies, instead of arbitrarily long. Indeed, let 
\begin{align}\label{eq:approximations}
 \forall \A\in \cA \ \forall X \in \cC \ \exists \A' \in \mathrm{conv}\,\cA' \  \exists X' \in \mathrm{conv}\,\cC':\quad | v(A,X) - v(A',X') | \leqslant \delta
\end{align}
where the $\mathrm{conv}$ operator denotes the convex hull. We get the following 
\begin{quote}
\textbf{Approximate Min-Max Theorem} If the condition \eqref{eq:approximations} holds, then the Weak Statement implies the Dream Statement is true with $\cA$ and $\cC$ replaced by $\cA'$ and $\cC'$.
\end{quote}
\subsection{Our contribution}\label{sec:main}
\textsc{Summary.}
This framework is well known (cf. \cite{Barak03,Reingold2008,Trevisan2009,Holenstein2005,VadhanZheng2013} to mention only some papers closely related to our cryptographic applications). What we offer, is a \emph{novel approximation technique}. Previous works used to find $\A'$ and $X'$ in convex hulls by a trivial Chernoff approximation argument. We observe that much better results are obtained with a carefully chosen \emph{convex approximation technique}. Indeed, it turns out that in many cases the quantity $| v(A,X) - v(A',X') | $ can be upper bounded by the \emph{H\"{o}lder Inequality} which involves moments of $A$ and $X$. These moments may be better estimated based on properties of the sets $\cA$ and $\cC$ which leads to quantitative improvements. We stress that the key component is \emph{the right choice of H\"{o}lder conjugates}, that is the exponents for the corresponding $L_p,L_q$ spaces.
\vspace{2mm}\newline\textsc{Advantages and Applications of our framework.} Using our technique we prove a whole bunch of results, reproving what is already known in a  more clear and concise way, improving quantitative bounds, or obtaining new results. Details are given in \Cref{sec:apps}.

\subsection{Related Works.}
The work of \cite{VadhanZheng2013} provides a tool to derive good bounds for certain sets $\cC$, in the uniform settings.
We stress that we consider only non-uniform adversaries here. In fact, our results can be probably made uniform by the use of constructive versions of auxiliary results on convex approximations we have applied (for example \cite{Donahue1997}). Anyway, uniform settings are not important for most of our applications like leakage-resilient crypto. While \cite{VadhanZheng2013} gives hard bounds, we provide \emph{a framework equipped with a different technique of handling $\cC$}. Our technique can exploit \emph{moment conditions}, which is impossible in \cite{VadhanZheng2013}. We stress that the crucial component of our technique is the 

\section{Applications}\label{sec:apps}
We briefly recall some basic notation and conventions. We say that two distributions $X_1,X_2$ are $(s,\epsilon)$-indistinguishable if for every $\A$ of size $s$ we have $|\E\A(X_1)-\E\A(X_2)| \leqslant \epsilon$.
\subsection{Impagliazzo Hardcore Lemma}
\textsc{Impagliazzo Hardcore Lemma.}  Suppose that are given a function $f:\{0,1\}^n\rightarrow \{0,1\}$ that is mildly hard to predict  by a class of circuits; for every circuit $\A$ from our class, $\A(x)$ and $f(x)$ agree on at most, say, a $0.99$ fraction of inputs $x$. This might happen when there is a set of noticeable size on which $f$ is extremely hard to predict, meaning that there is (almost) no advantage over a random guess. This set could be as big as a $0.02 = 2(1-0.99)$ fraction of input. Indeed, if $f$ cannot be guessed better than with probability $\frac{1}{2}$ on this set, then the probability that $\D$ agrees with $f$ is at most $ 0.02\cdot \frac{1}{2} + 0.98\cdot 1 = 0.99$.

Quite surprisingly, this intuitive characterization is true. The first such result was proved by Impagliazzo \cite{Impagliazzo95hard-coredistributions}, with a sub-optimal hardcore density. An improved version with the optimal density of the hardcore set was found by Holenstein \cite{Holenstein2005}. Below we present the best possible result due to Klivans and Servedio, the lower bound was given in \cite{Lu2007}.
\begin{theorem}[Optimal Unpredictability Hardcore Lemma \cite{Klivans2003}]\label{thm:unpredictability_hardcore_Holenstein}
Let $f:\{0,1\}^{n}\rightarrow \{0,1\}$ be $\epsilon$-unpredictable by circuits of size $s$ under a distribution $V$, that is
\begin{align}\label{eq:unpredictability_hardcore_if}
 \Pr_{x\leftarrow V}[\A(x) = f(x) ] \leqslant 1-\frac{\epsilon}{2},\quad \text{ for every $\A$ of size at most $s$.}
\end{align}
Then for any $\delta \in (0,1)$ there exists a event $E$ of probability $\epsilon$ such that $f$ is $1-\delta$ unpredictable under $V|E$ by circuits of size $s' = \bigOm {s \delta^2/\log(1/\epsilon) }$, that is
\begin{align}\label{eq:unpredictability_hardcore_then}
 \Pr_{x\leftarrow V|E}[\A(x) = f(x) ] \leqslant \frac{1+\delta}{2}, \quad \text{for every }\A \text{ of size at most } s'.
\end{align}
\end{theorem}
\textsc{Our Contribution.} We reprove \Cref{thm:unpredictability_hardcore_Holenstein} using the framework discussed in \Cref{sec:main}. Our approach has the following advantages over the related works:
\begin{enumerate}[(a)]
\item It is derived from the \emph{standard min-max theorem}. Previous proofs which achieve optimal parameters require involved iterative arguments \cite{Klivans2003,VadhanZheng2013}.
\item It is \emph{modular and much simpler} than all alternative proofs. Indeed, the argument of Holenstein is non-optimal and involved. Also the argument given by Vadhan and Zheng depends on a non-trivial trick attributed to Nissan and Levy (which improves the hardcore density from $\frac{\epsilon}{2}$ to $\epsilon$) and the machinery is much heavier. Our approach does not require this trick and follows the most intuitive strategy: show that there is a hardcore for every fixed adversary and then switch the order of quantifiers. 
\item We have identified \emph{the reason for non-optimality in previous proofs}. Some authors even suggested that it might be impossible to get the tight parameters using the standard min-max theorem \cite{VadhanZheng2013}. We show that this is not true. The problem is not with the standard min-max theorem but with an inadequate approximation argument in previous works, which do uniform approximation \cite{Holenstein2005}.
\end{enumerate}
A comparison is given below in \Cref{table:hardcore}.
\begin{table}[ht!]
\centering
\resizebox{0.95\textwidth}{!}{
\begin{tabular}{|c|l|c|l|}
\hline Author & Technique & Hardcore Density & Complexity Loss \\ \hline
\cite{Impagliazzo95hard-coredistributions} & boosting (constructive approx.) & $\Pr[E] = \frac{\epsilon}{2}$ & $O(\delta^{-2}\cdot\mathrm{poly}(1/\epsilon))$ \\ \hline
 \cite{Holenstein2005} & standard min-max + Hardcore Optimization & $\Pr[E] = \epsilon$ & $O(n\delta^{-2}) $\\ \hline
\cite{Lu2007} & complicated boosting (constructive approx.) & $\Pr[E] = \epsilon$ & $O(\log(1/\epsilon)\delta^{-2}) $\\ \hline
\cite{VadhanZheng2013} & complicated boosting (constructive approx.) & $\Pr[E] = \epsilon$ & $O(\log(1/\epsilon)\delta^{-2}) $\\ \hline
\textbf{this paper} & simple min-max + $L_p$-approx. & $\Pr[E] = \epsilon$ & $O(\log(1/\epsilon)\delta^{-2}) $\\ \hline
\end{tabular}
} 
\caption{Hardcore lemmas obtained by different techniques.}\label{table:hardcore}
\end{table}
\newline
\vspace{-15mm}\newline\textsc{A sketch of proof.} Assume without losing generality that $f:\{0,1\}^n\rightarrow \{-1,1\}$\footnote{We consider $\{-1,1\}$ outputs for technical convenience. Equivalently we could state the problem for $\{0,1\}$.}. Define the payoff $v$ as the unpredictability of $f$ by $A$ under $X$
\begin{align*}
v(A,X) \overset{\text{def}}{=} \Pr_{x\leftarrow X}[ f(x) = A(x) ] = \frac{1+\mathbb{E}_{x\leftarrow X} \A(x)\cdot f(x)}{2},
\end{align*}
and note that this definition makes sense also for circuits with real outputs.
Let the property set $\cC$ consists of conditional distributions of the form $X=V|E$ where $\Pr[E] \geqslant \epsilon$ and $E$ may vary\footnote{We can think of measures $M$ such that $M(\cdot)\leqslant \mathbf{P}_V(\cdot)$ and $\sum_x M(x) \geqslant \epsilon$. Every $X\in \cC$ can be written as $\mathbf{P}_X(\cdot) = M(\cdot)/\sum_{x}M(x)$ for one of these measures $M$.} ; note that $\cC$ is convex. Define $\cA$ as the set of real-valued\footnote{Following related works \cite{FullerReyzin12,Reingold2008} we use circuits with real outputs for technical reasons.} circuits of size $s$, and let $\cA'$ be the set of circuits of size $s'= \frac{s}{\delta^{-2}\log(1/\epsilon)}$. It is not hard to see that the assumption \eqref{eq:unpredictability_hardcore_if} implies  
\begin{proposition}[ Weak Statement ]\label{prop:unpredictability_hardcore_weak}
For every $\A\in\cA$ we have $v(X,A) \leqslant 0$ for some $X\in\cC$.
\end{proposition}
Now we analyze what happens when we replace $\cA$ by $\mathrm{conv}(\cA')$. We claim that
\begin{proposition}[Approximation Step]\label{prop:hardcore_approxstep}
For every $\A' \in \mathrm{conv}(\cA')$ we have $v(X,\A') \leqslant \delta$ for some $X\in\cC$.
\end{proposition}
To prove this, we show that the H\"{o}lder Inequality implies for $\A,\A'$ and $X\in \cC$ 
\begin{align*}
| v(X,\A) - v(X,\A')| & 
 \leqslant \frac{1}{2}\left( \mathbb{E}_{x\leftarrow V}\left( \frac{\mathbf{P}_{V|E}(x)}{\mathbf{P}_V(x)}\right)^{q} \right)^{\frac{1}{q}}\cdot \left( \mathbb{E}_{x\leftarrow V}\left| \A(x)-\A'(x) \right|^{p}  \right)^{\frac{1}{p}}
\end{align*}
for any $p,q\geqslant 1,\ \frac{1}{p}+\frac{1}{q} = 1$. Now we can argue that
\begin{enumerate}[(a)]
\item  $\left( \mathbb{E}_{x\leftarrow V}\left(\frac{\mathbf{P}_{V|E}(x)}{\mathbf{P}_V(x)}\right)^{q}\right)^{\frac{1}{q}}  \leqslant \epsilon^{-\frac{1}{p}}$ (by the extreme points technique).
\item $\left( \mathbb{E}_{x\leftarrow V}\left| \A(x)-\A'(x) \right|^{p}  \right)^{\frac{1}{p}} = O\left( \sqrt{\frac{p}{\ell}}\right)$ for some $\A$ which is of complexity $\ell$ relative to $\cA'$\footnote{That is, $\A$ is a convex combination of $\ell$ members of $\A'$} (by standard facts on convex-approximation \cite{Docampo1997}).
\end{enumerate}
Setting $\ell = \delta^{-2}\log(1/\epsilon)$ (so that $A\in \cA)$, taking $X = V|E$ which corresponds to $\A'$ according to \Cref{prop:unpredictability_hardcore_weak}, setting $p=2\log(1/\epsilon)$ and putting this all together we get \Cref{prop:hardcore_approxstep}. This implies the following statement
\begin{proposition}[ Strong Statement ]
For some $X\in\cC$ we have $v(X,A) \leqslant \delta$ for every $\A\in\cA'$.
\end{proposition}
which proves \Cref{thm:unpredictability_hardcore_Holenstein} ($|v(X,A)| \leqslant \delta$ follows by considering $\cA'$ closed under complements).

\subsection{A (new) optimal hardcore lemma for metric pseudoentropy and applications to transformations.}

Pseudoentropy notions extend classical information-theoretic entropy notions into computational settings. The following most widely used entropy notions capture what it means to be ``computationally close'' to a high entropy distribution.
\begin{definition}[HILL Pseudoentropy \cite{HILL99}]
Let $Y$ be a distribution with the following property: there exists $Y'$ of min-entropy at least $k$ such that for every $\A$ of size at most $s$ we have $|\E\A(Y)- \E Y')|\leqslant \epsilon$. Then we say that $X$ has $k$ bits of HILL entropy of quality $(s,\epsilon)$ and denote by $\mathbf{H}_{s,\epsilon}^{\textup{HILL}}(Y)\geqslant k$.
\end{definition}
\begin{definition}[Metric Pseudoentropy \cite{Barak03}]
Let $Y$ be a distribution with the following property: for every $\A$ of size at most $s$ there exists $Y'$ of min-entropy at least $k$ such that we have $|\E\A(Y)- \E Y')|\leqslant \epsilon$. Then we say that $X$ has $k$ bits of metric entropy of quality $(s,\epsilon)$ and denote by $\mathbf{H}_{s,\epsilon}^{\textup{Metric}}(Y)\geqslant k$.
\end{definition}
Pseudoentropy is an important research area, with applications in deterministic encryption, memory delegation \cite{Chung2011}, pseudorandom generators \cite{HILL99,VadhanZheng2013}. Metric Pseudoentropy is much easier to deal with, and fortunately can be converted into HILL entropy with some loss in quality parameters $(s,\epsilon)$.
\vspace{2mm}\newline\textsc{Our contribution.}
The following results shows that any distribution with metric pseudoentropy of `moderate'' quality has a kernel of HILL entropy with ``strong'' quality. We also conclude the optimal Metric-HILL transformation.
\begin{theorem}[A HILL-pseudoentropy hardcore for metric pseudoentropy]\label{thm:metric_hardcore}
Suppose that $\mathbf{H}_{s,\epsilon}^{\textup{Metric}}(Y)\geqslant n-\Delta$, for some $Y\in\{0,1\}^n$. Then there is an event $E$, of probability $1-\epsilon$ such that $\mathbf{H}_{s',\delta}^{\textup{HILL}}(Y|E)\geqslant n-\Delta$ with $s'=\Omega( s\delta^2/(\Delta+1))$ for every $\delta$. In particular, $\mathbf{H}_{s',\epsilon+\delta}^{\textup{HILL}}(Y)\geqslant n-\Delta$
\end{theorem}
One possible application of this fact is amplifying hardness of pseudoentropy with poor quality. Imagine that we have many independent samples $X_1,X_2,\ldots, X_n$ from a distribution with a substantial entropy amount ($\Delta \ll n$) but of weak advantage $\epsilon = 0.99$. We can use the result above to show that pseudoentropy in $X_1,X_2,\ldots, X_n$ is roughly $(1-0.99)(n-\Delta)$ with good quality (see \cite{DBLP:conf/icits/Skorski15a} for more details).
Below we briefly compare this result with related works.
\begin{enumerate}[(a)]
\item Our result is \emph{far stronger than the classical result due to Barak et al. \cite{Barak03}}  about the transformation. Not only we replace the factor $n$ by $\Delta$, but also show the existence of a hardcore in the intermediate step.
\item This result \emph{unifies and improves our recent results \cite{DBLP:conf/icits/Skorski15,DBLP:conf/icits/Skorski15a}}. The corollary $\mathbf{H}_{s',\epsilon+\delta}^{\textup{HILL}}(Y)\geqslant n-\Delta$ was the same (and optimal) but the hardcore $E$ was found with worse complexity $s' = \Omega(s\cdot \delta^2/n)$. 
\item Our result \emph{explains the nature of the Metric-HILL transformation}. The HILL pseudoentropy hardcore is an intermediate step in going from Metric pseudoentropy to HILL pseudoentropy. 
\end{enumerate}
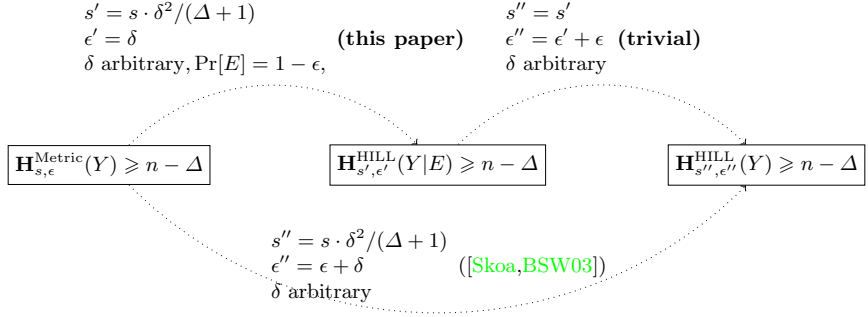
\begin{figure}[ht!, scale=0.8]
\centering
\resizebox{0.95\textwidth}{!}{
\begin{tikzpicture}
\draw (0,0) node[draw] (a) {$\mathbf{H}_{s,\epsilon}^{\textup{Metric}}(Y)\geqslant n-\Delta$};
\draw (5,0) node[draw] (b) {$\mathbf{H}_{s',\epsilon'}^{\textup{HILL}}(Y|E)\geqslant n-\Delta$};
\draw (10,0) node[draw] (c) {$\mathbf{H}_{s'',\epsilon''}^{\textup{HILL}}(Y)\geqslant n-\Delta$};
\path[->,dotted] (a) edge [bend right = -45] node[above] {$\begin{array}{l} s'=s\cdot \delta^2/(\Delta+1) \\ \epsilon'=\delta \\\delta \text{ arbitrary},  \Pr[E]=1-\epsilon,  \end{array}$ \textbf{(this paper)}} (b);
\path[->,dotted] (b) edge [bend right = -45] node[above] {$\begin{array}{l} s''=s' \\ \epsilon''=\epsilon'+\epsilon \\ \delta \text{ arbitrary} \end{array}$ \textbf{(trivial)}} (c);
\path[->,dotted] (a) edge [bend left = -45] node[above] {$\begin{array}{l} s''=s\cdot \delta^2/(\Delta+1) \\ \epsilon''=\epsilon+\delta \\ \delta \text{ arbitrary } \end{array}$ (\cite{DBLP:conf/icits/Skorski15,Barak03}) } (c);
\end{tikzpicture}
}
\caption{The Metric-to-HILL pseudoentropy transformation.}\label{fig:Metric-to-HILL}
\end{figure}
Our result is illustrated in \Cref{fig:Metric-to-HILL}. The parameters are optimal (see \cite{DBLP:conf/icits/Skorski15a}).
\newline
\vspace{-1mm}\newline\textsc{A sketch of proof.}
Let $\cA$ be the set of real-valued circuits of size $s$ and let $\cA'$ be the set of circuits of size $s'=s\delta^2/(\Delta+1)$.
Let $\mathcal{C}$ consists of the conditional distributions $X$ of the form $X'|E$, where $\Pr[E] \geqslant 1-\epsilon$ and $\mathbf{H}_{\infty}(X') \geqslant n-\Delta$; this set is convex. The payoff is defined as $
 v(X,A) \overset{def}{=} \mathbb{E}\A(Y) - \mathbb{E}\A(X) 
$. It is easy to see\footnote{This is trivial for boolean $\A$ and somewhat more tricky for real-valued $\A$. A short proof is given implicitly in \cite{FullerReyzin12}} that we have
\begin{proposition}[ Weak Statement ]\label{prop:metric_weak}
$ \forall \A\in \cA\ \exists X\in \cC \quad  v(X,A) \leqslant 0$.
\end{proposition}
Now we analyze what happens when $\cA$ is replaced by $\mathrm{conv}(\cA')$.
\begin{proposition}[Approximation Step]\label{prop:hardcore_approxstep}
For every $\A' \in \mathrm{conv}(\cA')$ we have $v(X,\A') \leqslant \delta$ for some $X\in\cC$.
\end{proposition}
To prove this, by the H\"{o}lder Inequality for any $\A,\A'$ and $X\in \cC$ we show   
\begin{align*}
| v(X,\A) - v(X,\A')| & 
 \leqslant \left( \mathbb{E}_{x\leftarrow U}\left( 2^{n}\mathbf{P}_{Y|E}(x)\right)^{q} \right)^{\frac{1}{q}}\cdot \left( \mathbb{E}_{x\leftarrow U}\left| \A(x)-\A'(x) \right|^{p}  \right)^{\frac{1}{p}}
\end{align*}
for any $p,q\geqslant 1,\ \frac{1}{p}+\frac{1}{q} = 1$ and the uniform distribution $U$. Now we argue that
\begin{enumerate}[(a)]
\item  $\left( \mathbb{E}_{x\leftarrow U}\left( 2^{n}\mathbf{P}_{Y|E}(x)\right)^{q} \right)^{\frac{1}{q}} \leqslant 2^{\frac{\Delta}{p}}$ (by the extreme points technique).
\item $\left( \mathbb{E}_{x\leftarrow V}\left| \A(x)-\A'(x) \right|^{p}  \right)^{\frac{1}{p}} = O\left( \sqrt{\frac{p}{\ell}}\right)$ for some $\A$ which is of complexity $\ell$ relative to $\cA'$ (by standard facts on convex-approximation \cite{Docampo1997}).
\end{enumerate}
Setting $\ell = \delta^{-2}(\Delta+1)$ (so that $A\in \cA)$, taking $X = X'|E$ which corresponds to $\A'$ according to \Cref{prop:unpredictability_hardcore_weak}, setting $p=\Delta+1$ and putting this all together we get \Cref{prop:hardcore_approxstep}. This implies the following statement
\begin{proposition}[ Strong Statement ] $\exists X\in \cC\ \forall \A\in \cA'\quad  v(X,A) \leqslant \delta$.
\end{proposition}
This directly implies \Cref{thm:metric_hardcore} (as before, we consider $\A'$ closed under complements). More details can be found in \Cref{app:thm:metric_hardcore}.

\subsection{A (fixed) construction of a simulator for auxiliary inputs.}
In~\cite{JetchevP14} there is a theorem, which says that any short information $Z$ about $X$ can be efficienly simulated from $X$, Below we state the corrected version~\cite{Pietrzak15_private}.
\begin{theorem}[Simulating auxiliary inputs, flaws fixed]\label{thm:aux_inputs}
For any random variable $X\in\{0,1\}^n$, any correlated $Z\in\{0,1\}^\lambda$ and every choice of parameters $(\epsilon,s)$ there is a randomized function $\SimS:\{0,1\}^n\rightarrow \{0,1\}^\lambda$ of complexity $\bigO{s\cdot 2^{4\lambda}\epsilon^{-4}}$ such that $Z$ and $\Sim{X}$ are $(\epsilon,s)$-indistinguishable given $X$.
\end{theorem}
This result is the key component in the simplified analysis of the EUROCRYPT'09 stream cipher construction. Using \Cref{thm:aux_inputs}, as described in~\cite{JetchevP14}, one proves the resilience of the cipher (assuming bounded leakage in every round) and if the underlying weak PRF is $(s,\epsilon)$-secure against two queries on random inputs. The cipher security $(s',\epsilon')$ is related to $(s,\epsilon)$ by a polynomial loss in $\epsilon$. 
\vspace{2mm}\newline\textsc{Our contribution.} We describe a flaw in the proof and improve the corrected bound by a significant super polynomial factor. Below we briefly describe the significance  of our result
\begin{enumerate}[(a)]
\item \emph{Discovered flaws} in the recent (TCC'14) analysis of the EUROCRYPT'09 stream cipher. The alternative bounds seem correct but are much weaker. In particular, we get \emph{no meaningful security with the AES} used as a weak PRF in this construction\footnote{The final bounds on the cipher security depends on the simulator complexity and are given by $\epsilon' = O\left( \sqrt{2^{\lambda}\epsilon}\right)$ and $s'=s\cdot 2^{-4\lambda}\epsilon'^{4}$. We can't prove then even very weak security $\epsilon' = 2^{-32}$ having $\lambda=10$ bits of leakage!}. This raises the problem of \emph{whether the cipher built on AES is secure or not}. We would need a simulator with a loss of only $O(\epsilon^{-2})$ not $\epsilon^{-4}$ in complexity.
\item A \emph{simpler} construction based on the min-max theorem. Based on the framework in \Cref{sec:main} we give an alternative proof achieving the simulator complexity of $O\left( s\cdot 2^{2\lambda}\epsilon^{-4}\right)$). The gain of $2^{2\lambda}$ over the original approach, which is a power of $\epsilon$ for recommended values of parameters~\cite{JetchevP14}, comes from the use of convex approximation techniques. Our proof is considerably simpler and quantitatively better than in \cite{JetchevP14} (in particular we don't need to use the min-max theorem twice depending on what is the value of the game). Also, it is much simpler than the alternative approach of Vadhan and Zheng \cite{VadhanZheng2013}, yet yields comparable results for small leakages (see \Cref{table:stream_ciphers_strength}). 
\item A \emph{clear bound on the security level}, in terms of the time-success ratio.
We derive a clear formula which shows what fraction of the security of the original weak PRF is transformed into security of the stream cipher. This analysis shows that we are \emph{far from good and provable secure} leakage-resilient stream ciphers as we lose over $\frac{5}{6}$ of original security. For more details, see \Cref{table:stream_ciphers_strength}.
\end{enumerate}
In \Cref{table:stream_ciphers_strength} we compare the strength of the simulator theorems in terms of implied security for this construction. To our knowledge, this is the first analysis of the time-success ratio for this technique. For more details we refer to \Cref{app:leakage-resilient}.
\begin{table}[ht!]
\centering
\resizebox{0.95\textwidth}{!}{
\begin{tabular}{|c|c|l|c|}
\hline Author & Technique & Simulator Complexity & Implied Security \\ \hline
\cite{JetchevP14} & Standard Min-Max + $L_{\infty}$-approx. & $s_{h} = s\cdot 2^{4\lambda}\epsilon^{-4} $ & $k' = \frac{k}{6}-\frac{5}{6}\lambda$ \\ \hline
\cite{VadhanZheng2013} & Complicated Boosting & $s_{h} = s\cdot 2^{\lambda}\epsilon^{-2} + 2^{\lambda}\epsilon^{-4}$ & $k' = \frac{k}{6}-\frac{1}{3}\lambda$ \\ \hline
\textbf{this paper} & Standard Min-Max + $L_p$-approx. & $s_{h} = s\cdot 2^{2\lambda}\epsilon^{-4} $ & $k' = \frac{k}{6}-\frac{1}{2}\lambda$ \\ \hline
\end{tabular}
}
\caption{Security of the EUROCRYPT'09 stream cipher instantiated with a wPRF having $2^{k}$ keys and $\lambda$ bits of leakage, obtained from different simulator results. Every attacker of size $s$ succeeds with prob. at most $s/2^{k'}$}
\label{table:stream_ciphers_strength}.
\end{table}
\vspace{2mm}\newline\textsc{More on the flaws.}
In the claimed better bound $\bigO{s\cdot 2^{3\lambda}\epsilon^{-2}}$ there is a mistake on page 18 (eprint version), when the authors enforce a signed measure to be a probability measure by a mass shifting argument. The number $M$ defined there is in fact a function of $x$ and is hard to compute, whereas the original proof assumes that this is a constant independent of $x$. In the alternative bound $\bigO{s\cdot 2^{3\lambda}\epsilon^{-2} }$ a fixable flaw is a missing factor of $2^{\lambda}$ in the complexity (page 16 in the eprint version), which is because what is constructed in the proof is only a probability mass function, not yet a sampler~\cite{Pietrzak15_private}.
\vspace{2mm}\newline\textsc{A sketch of the proof.} Let $\cA$ be the set of real-valued circuits of size $s$ and let $\cA'$ be the set of circuits of size $s'=s\cdot 2^{-2\lambda}\epsilon^{2}$.
Let $\mathcal{C'}$ consists of the distributions of the form $X,h(X)$, where $h$ is computable in size $ s\cdot 2^{\lambda} $; this set is \emph{not} convex. Let $\cC$ be the set of all circuits of size $s\cdot 2^{2\lambda}\epsilon^{-2}$.
The payoff is defined as $
 v(h,A) \overset{def}{=} \mathbb{E}\A(X,h(X)) - \mathbb{E}\A(X,Z) 
$. It is easy to see that we have
\begin{proposition}[ Weak Statement ]\label{prop:simulating_weak}
$ \forall \A \in \cA\ \exists h'\in \cC' \quad  v(h',A) \leqslant 0$.
\end{proposition}
Indeed, consider $h_{\A}^{+}$ which for every $x$ outputs this value $z$ for which $\A(x,z) = \max\A(x,\cdot)$ and $h_{\A}^{-}$ which for every $x$ outputs this value $z$ for which $\A(x,z) = \min\A(x,\cdot)$. Both are of complexity $O\left(2^{\lambda}\right)$. Since we have $\E\A(X,h^{-}(X)) \leqslant \E\A(X,Z)$ and $ \E\A(X,Z) \leqslant \E\A(X,h^{+}(X))$, setting $h'$ to be a distribution over $h^{+}$ and $h^{-}$ that is $\Pr[h'(x)=z] =\theta \cdot \Pr[h^{-}(x)=z] + (1-\theta) \cdot \Pr[h^{+}(x)=z]$, we get $v(h',\A)=0$ with some $\theta$. In the next step we replace $\cA'$ by $\mathrm{conv}(\cA')$.
\begin{proposition}[ Approximation 1 ]\label{prop:simulating_approx1}
$\forall A\in \mathrm{conv} \cA' \ \exists h'\in \cC':\quad v(h',\A') \leqslant \epsilon$.
\end{proposition}
This follows from the standard Chernoff Bound approximation argument\footnote{$A$ can be viewed as a distribution on $\cA'$ we simply pick $\ell$ independent samples $\{\A_i\}_i$ and try to find an approximator of the form $\A'=\frac{1}{\ell}\sum_{i=1}^{\ell} \A_{i} $. It deviates by more than $\epsilon$ at $(x,z)$ with probability $\exp(-2\ell\epsilon^2)$. We combine this with the union bound.} as
\begin{align*}
 | v(h',A) - v(h',A')| =| \E\A(X,h'(X))-\E\A'(X,h'(X)) | \leqslant  \sup_{x,z} | \A(x,z)-\A'(x,z) |.
\end{align*}
Now we replace $\cC'$ by $\mathrm{conv}\ \cC'$. Here a more delicate approximation is required.
\begin{proposition}[ Approximation 2]\label{prop:simulating_approx2} For every $A$ and every $h'\in \mathrm{conv}\ \cC'$ there exists $h\in \cC$ such that $ v(h,\A) \leqslant  v(h',\A) + \epsilon$.
\end{proposition}
This follows because by the H\"{o}lder Inequality applied to $p=q=2$ we obtain
\begin{align*}
| \E\A(X,h'(X))-\E\A(X,h(X)) | &\leqslant 2^{\frac{\lambda}{2}}\cdot\left( \E_{x\sim X}\sum_{z}\left| \mathbf{P}_{x,h(x)}(x,z)-\mathbf{P}_{x,h'(x)(x,z)}\right|^2 \right)^{\frac{1}{2}},
\end{align*}
and by the standard results on convex approximation~\cite{Donahue1997} the second factor is at most $\ell^{-\frac{1}{2}}$ for some $h$ of complexity  $\ell$ with respect to $\cC'$. We put $\ell = 2^{\lambda}\epsilon^{-2}$. From the proven propositions we obtain the final result.
\begin{proposition}[ Strong Statement]\label{prop:simulating_approx2} $\exists h\in \cC\ \forall \A\in \cA' \quad v(h,\A) \leqslant  2\epsilon$.
\end{proposition}

\subsection{More Applications}
For more applications we refer an interested reader to \Cref{sec:more_applications}. They include the optimal Dense Model Theorem, a better auxiliary input simulator for bounded-variance adversaries (new), and a proof that every high-conditional entropy source can be efficiently simulated (new, extending \cite{Trevisan2009}).

\bibliographystyle{amsalpha}
\bibliography{./citations}

\newpage

\appendix

\section{More Applications}\label{sec:more_applications}.

\subsection{Dense Model Theorem}

Given a pair of two distributions $W$ and $V$ over the same finite domain we say that $W$ is $\delta$-dense in $V$ if and only if $\Pr[W=x]\leqslant\Pr[V=x]/\delta$\footnote{The term ``$\delta$-dense'' comes from the fact that $V$ can be written as a convex combination of $W$ with weight $\delta$ and some other distribution with weight $1-\delta$}.
The efficient version of the famous dense model theorem specialized to the boolean case, can be formulated as follows:
\begin{theorem}[Dense Model Theorem.]\label{thm:DMT} Let $\mathcal{D'}$ be a class of $n$-bit boolean functions, $R$ be uniform over $\{0,1\}^n$, $X$ be an $n$-bit random variable and let $X'$ be $\delta$-dense in $X$. If $X$ and $R$ are $(\mathcal{D},\epsilon)$-indistinguishable then there exists a distribution $R'$ which is $\delta$-dense in $R$ such that $X'$ and $R'$ are $(\mathcal{D}',\epsilon')$-indistinguishable, where $\epsilon' = (\epsilon/\delta)^{O(1)}$ and $\mathcal{D}$ consists of all functions of the form $g(D_1,\ldots,D_{\ell})$ where $D_i \in \mathcal{D}'$, $\ell = \mathrm{poly}(1/\delta,1/\epsilon)$ and $g$ is some function.
\end{theorem}
Using our framework we can reprove the Dense Model Theorem with optimal parameters due to Zhang \cite{Zhang2011}. The proof is very similar to the one in \Cref{thm:metric_hardcore} so we omit the details; we note that we need the H\"{o}lder Inequality with $p = 2\log(1/\delta)$. A similar technique appears in \cite{DBLP:conf/icits/Skorski15}, though we do note use Metric pseudoentropy here.
\begin{corollary}
Dense Model Theorem (\Cref{thm:DMT}) holds with $\epsilon' = O(\epsilon/\delta)$, $g$ being a linear threshold and $\ell = O(\log(1/\delta)  / (\epsilon/\delta)^2$. 
\end{corollary}
\begin{table}
\centering
\resizebox{0.95\textwidth}{!}{
\begin{tabular}{| l | c | c | c | c|}
\hline
 Author & Technique & Function $g$ &  $\ell$ as complexity of $\mathcal{D}'$ w.r.t $\mathcal{D}$  & $\epsilon'$ vs $\epsilon$ \\
\hline
Tao and Ziegler & Complicated & Inefficient & $\ell=\poly(1/(\epsilon/\delta),\log(1/\delta))$ & $\epsilon' = O(\epsilon/\delta)$ \\ \hline
 \cite{Reingold2008} & Min-Max Theorem & Linear threshold & $\poly(1/(\epsilon/\delta),\log(1/\delta))$ & $\epsilon' = O(\epsilon/\delta)$ \\ \hline
\cite{FullerReyzin12}, \cite{Dziembowski2008} & Metric Entropy & Linear threshold & $\ell= O(n  / (\epsilon/\delta)^2 ) $ & $\epsilon' = O( \epsilon/\delta)$ \\ \hline
\cite{Zhang2011} & Boosting & Linear threshold & $\ell= O(\log(1/\delta)  / (\epsilon/\delta)^2 $ & $\epsilon' = O( \epsilon/\delta)$ \\ \hline
\ \textbf{This paper} & Standard Min-Max + $L_p$-approx & Linear threshold & $\ell= O(\log(1/\delta)  / (\epsilon/\delta)^2 $ & $\epsilon' = O( \epsilon/\delta)$ \\ \hline
\end{tabular} 
}
\caption{Different versions of the Dense Model Theorem}\label{table:comparison}
\end{table}

\subsection{Simulating auxiliary inputs against bounded-variance distinguishers}
An interesting result is obtained when working more carefully in the proof of \Cref{thm:aux_inputs}. Namely, imposing additional restriction on the second moment of test functions we obtain a refined bound
\begin{theorem}[Simulating auxiliary inputs against bounded-variance distinguishers]\label{thm:aux_inputs_variance}
For any random variable $X\in\{0,1\}^n$, any correlated $Z\in\{0,1\}^\lambda$, any class $\cA$ be of functions $\A:\{0,1\}^n\times\{0,1\}^m\rightarrow [0,1]$ such that $\forall\A\in\cA:\ \E_{X} \Var\A(x,U) \leqslant \sigma^2$, and every $\epsilon$ there is a randomized function $\SimS:\{0,1\}^n\rightarrow \{0,1\}^\lambda$ of complexity $\bigO{s\cdot 2^{4\lambda}\textcolor{red}{\sigma}\epsilon^{-4}}$ relative to $\cA$ such that $Z$ and $\Sim{X}$ are $\epsilon$-indistinguishable given $X$ by functions $\cA$.
\end{theorem}
This result is interesting in the context of recent improvements in key derivation, so called square security, where the second moment condition is widely used.

\subsection{Simulating High Entropy Distribution with Auxiliary Information}
From our framework we derive the following result, which is the extension of the theorem in \cite{Trevisan2009} into a conditional case (in the presence of auxiliary information). We stress that this result \emph{cannot} be derived from the techniques used in \cite{Trevisan2009}, because this approach will not preserve the same marginal distribution $Z$, when applied in the conditional setting.

\begin{theorem}[High conditional min-entropy is simulatable]\label{thm:simulating_high_entropy}
Let $X\in \{0,1\}^n$ and $Z\in\{0,1\}^m$ be correlated random variables and $\mathbf{H}_{\infty}(X|Z) = n-\Delta$.
 Then there exists a distribution $Y,Z$ such that
\begin{enumerate}[(a)]
\item There is a circuit $\mathsf{Sim}$ of complexity $\bigO{n(n+m)2^{2\Delta}\epsilon^{-5}}$ and such that $\mathsf{Sim}(Z) \overset{d}{=} Y$
\item $(X,Z)$ and $(Y,Z)$ are $(s,\epsilon)$-indistinguishable
\item We have $\mathbf{H}_{\infty}(Y|Z) \geqslant n-\Delta-6$.
\end{enumerate}
\end{theorem} 
Here we show only how to simulate given a fixed distinguisher. The rest of the proof follows by
the use of a convex approximation argument (with $p=q=2$) and allows us to save a factor of $2^{2\Delta}$ in the simulator complexity.

\begin{proof}[Proof of the weak statement]
Let $\Delta = n-k$. By replacing $\epsilon$ with $2\epsilon$ we can assume that $\cD = \sum_{i=1}^{j}\alpha_i \cD_i$ where $\alpha_i = 1-(i-1)\epsilon$ for $i=1,\ldots, \lceil 1/\epsilon \rceil$ and $\cD_i$ are boolean such that $\mathbf{1}=\sum_{i}\cD_i$. Define 
\begin{align}
d(i) = \Pr[\D(U)\geqslant \alpha_i ].
\end{align}
and let $M$ be the smallest number $i$ such that $d(i) \geqslant 2^{-\Delta}$. Note that if we didn't care about computational efficiency then the best answer would be
\begin{equation}
 Y^{+} \overset{d}{=} \frac{d(M-1)}{2^{-\Delta}}\cdot U_{\cD_1+\ldots + \cD_{M-1}} +\frac{2^{-\Delta}-d(M-1)}{2^{-\Delta}}\cdot U_{\cD_M}
\end{equation}
because then
\begin{align}
\mathbf{E}\D(Y^{+}) & = \frac{\sum_{i=1}^{M-1}\alpha_i|\D_i| + \left(2^k-\sum_{i=1}^{M-1}\alpha_i|\D_i| \right)\alpha_M}{2^k} \nonumber \\
& = \max_{Y:\ \mathbf{H}_{\infty}(Y)\geqslant k}\mathbf{E}\D(Y)
\end{align}
The approach we chose is quite obvious - we efficiently approximate the distribution $Y^{+}$. For any $i$, sample $x_1,\ldots,x_{\ell}$ where $\ell > 2^{\Delta}n\log(1/\epsilon)/\epsilon$ and let
\begin{align}
 \tilde{d}(i) = \ell^{-1}\sum\limits_{j=1}^{\ell} \mathbf{1}_{\{ \D(x_i) \geqslant \alpha_i \}}
\end{align}
Now let $M'$ be the smallest number such that $\tilde{d}(M') > \frac{3}{4}\cdot 2^{-\Delta}$. Note that that $M'$ is well defined with probability $1-2^{-n}$, and then we have
\begin{align}
 \tilde{d}(M'-1) < \frac{3}{4}\cdot 2^{-\Delta} < \tilde{d}(M')
\end{align}
Now we define $Y$ as follows:
\begin{align}
 Y \overset{d}{=} \left\{ \begin{array}{rl}
  \frac{\tilde{d}(M'-1)}{2^{-\Delta}}\cdot\widetilde{U}_{D_1+\ldots+D_{M'-1} } + \left(1-\frac{\tilde{d}(M'-1)}{2^{-\Delta}}\right)\cdot \widetilde{U}_{\cD_{M'}} &
  2^{-\Delta}\epsilon  < \tilde{d}(M'-1) < 2^{-\Delta}/16
    \\
  \widetilde{U}_{D_1+\ldots+D_{M'-1} }, & 2^{-\Delta}/16 < \tilde{d}(M'-1)  \\
  \widetilde{U}_{\cD_{M'}}, &  2^{-\Delta}\epsilon > \tilde{d}(M'-1) 
 \end{array} \right.
\end{align}
Observe that if $d(i) < 2^{-\Delta}/4$ then with probability $1-2^{-n}$ we get
$\tilde{d}(i) < 2^{-\Delta}/2$. Thus, the probability that $d(M') < 2^{-\Delta}/4$ and $\tilde{d}(M') > 2^{-\Delta}/2$ is at most $2^{-n}\log(1/\epsilon)$ and we can assume that $d(M')>2^{-\Delta}/4$. Similarly, if $d(i) > 2^{-\Delta}$ then with probability $1-2^{-n}$ we have $\tilde{d}(i) > \frac{3}{4}\cdot 2^{-\Delta}$ which means $M' \leqslant i$. Therefore, with probability $1-2^{-n}\log(1/\epsilon)$ we can assume that $d(M'-1) < 2^{-\Delta}$. Now we split the analysis into the following cases
\begin{enumerate}[(a)]
\item $\tilde{d}(M'-1) < 2^{-\Delta}\epsilon$ and $d(M'-1) < 2\cdot 2^{-\Delta}\epsilon$.  Since
$|\D_{M'}|=2^{n}(d(M')-d(M'-1))\geqslant 2^{n-\Delta}/8$, we see that $U_{\cD_{M'}}$ is samplable in time $\mathcal{O}\left(2^{\Delta}\log(1/\epsilon)\right)$ and that $\mathbf{H}_{\infty}(U_{\cD_{M'}})\geqslant k-3$. Note that
\begin{align}
\mathbf{E}\D(Y^{+}) & = \mathbf{E}\D(Y^{+})\mathbf{1}_{\D(Y^{+}) \geqslant \alpha_{M'-1}} + 
\mathbf{E}\D(Y^{+})\mathbf{1}_{\D(Y^{+}) \leqslant \alpha_{M'}}  \nonumber \\
& \leqslant 2\epsilon + \alpha_{M'} \nonumber \\
& \leqslant 3\epsilon + \mathbf{E}\D(Y)
\end{align}
\item $ \tilde{d}(M'-1) > 2^{-\Delta}/16$ and $2^{-\Delta} > d(M'-1) > 2^{-\Delta}/32$. Then we have $|\D_1|+\ldots + |\D_{M'-1}|\geqslant 2^{n-\Delta-5}$ and thus $\mathbf{H}_{\infty}(\widetilde{U}_{\cD_1+\ldots+\cD_{M'-1} }) \geqslant n-\Delta-5$ and $\widetilde{U}_{\cD_1+\ldots+\cD_{M'-1} }$ is samplable in time $\mathcal{O}\left(2^{\Delta}\log(1/\epsilon)\right)$. Since $|\D_1|+\ldots + |\D_{M'-1}|\leqslant 2^{n-\Delta}$, we have
\begin{align}
 \mathbf{E}\D(Y^{+}) & \leqslant \mathbf{E}\D(U_{ \cD_1+\ldots + \cD_{M'-1}}) \nonumber \\
 & \leqslant \mathbf{E}\D(\widetilde{U}_{ \cD_1+\ldots+\cD_{M'-1} }) + \epsilon
\end{align}
\item $2^{-\Delta}\epsilon  < \tilde{d}(M'-1) < 2^{-\Delta}/16$ and $2^{-\Delta}\epsilon/2 < d(M'-1) < 2^{-\Delta}/8$ and $\tilde{d}(M'-1) \leqslant 2d(M'-1)$.
 We have $|\D_1|+\ldots + |\D_{M'-1}|=2^{n}d(M'-1)>2^{n-\Delta}\epsilon/2$ and $|D_{M'}| = 2^{n}(d(M')-d(M'-1))\geqslant 2^{n-\Delta}/8$, therefore $Y$ is samplable in time $\mathcal{O}\left(2^{\Delta}\log(1/\epsilon)/\epsilon \right)$. Moreover, we have $\mathbf{H}_{\infty}(\widetilde{U}_{\cD_1+\ldots+\cD_{M'-1}}) \geqslant \log(|\D_1|+\ldots+|\D_{M'-1}|)$ and $\mathbf{H}_{\infty}(\widetilde{U}_{\cD_{M'}}) \geqslant \log |D_{M'}|$. Hence $\mathbf{H}_{\infty}(\widetilde{U}_{\cD_1+\ldots+\cD_{M'-1}})  \geqslant n+\log d(M'-1)$ and $\mathbf{H}_{\infty}(\widetilde{U}_{\cD_{M'}})\geqslant n-\Delta-3$ and
\begin{align}
\Pr[Y=x] & \leqslant \frac{\tilde{d}(M'-1)}{d(M'-1)}\cdot 2^{-n+\Delta} + 2^{-n+\Delta+3} \nonumber \\
& \leqslant 2^{-n+\Delta+4}
\end{align}
\end{enumerate}
Suppose now that $d(M'-1)< 2^{-\Delta}\epsilon/2$. Then, by the Chernoff Bound with probability $1-2^{-n}$ we have $\tilde{d}(M'-1) < 2^{-\Delta}\epsilon/2+d(M'-1)<2^{-\Delta}\epsilon$ and we are in case (a). If $2^{-\Delta}\epsilon/2 < d(M'-1)< 2^{-\Delta}/32$ then with probability $1-2^{-n}$ we have $\frac{1}{2} < \frac{\tilde{d}(M'-1)}{d(M'-1)} < 2$ and it is easy to check that we can be either in (a) or in (c), depending on $\tilde{d}(M'-1)$. If $2^{-\Delta}/32<d(M'-1)<2^{-\Delta}/8$ then with probability $1-2^{-n}$ we are either in (c) or in (b). If $2^{-\Delta}/8<d(M'-1)<2^{-\Delta}$ then with probability $1-2^{-n}$ we can be only in (b). 
\end{proof}

\section{Convex Approximation Rates}\label{sec:convex_approximation}

We use the following fact on convex approximation rates.
\begin{lemma}[Convex approximation in $L^{p}$ spaces \cite{Donahue1997}]\label{lemma:lp_approximation}
Let $\mathcal{X}$ be a finite domain, $\nu$ be a distribution on $\mathcal{X}$. Fix a number $1\leqslant p < +\infty$ and for any function $f$ on $\mathcal{X}$ define $\left\| f\right\|_{p} = \left(\E_{x\leftarrow \nu} | f(x)|^{p}\right)^{\frac{1}{p}}$. Let $\mathcal{G}$ be any set of real functions on $\mathcal{X}$, let $\overline{g}$ be a convex combinations of functions from $\mathcal{G}$ and $K>0$ be such that for all $g\in \mathcal{G}$ we have $\|\overline{g}-g\|_p\leqslant K$. Then for any $\ell > 0$ there exists a convex combination $g' = \sum_{i=1}^{\ell} \alpha_i g_i$ of functions $g_1,\ldots,g_k \in \mathcal{G}$ such that
\begin{align*}
 \|  \overline{g}- g'\|_p \leqslant \frac{K C_p}{\ell^{1-\frac{1}{t}}}
\end{align*}
where $t=\min(2,p)$ and $C_p=1$ if $1\leqslant p\leqslant 2$, $C_p = \sqrt{2}[ \Gamma( (p+1)/2) / \sqrt{\pi} ]^{1/p}$ for $2<p<+\infty$. 
\end{lemma}

\section{Proof of \Cref{thm:metric_hardcore}}\label{app:thm:metric_hardcore}
To finish the proof it remains to justify the estimates
\begin{align}\label{eq:thm:metric_hardcore_1}
\left( \mathbb{E}_{x\leftarrow U}\left( 2^{n}\mathbf{P}_{Y|E}(x)\right)^{q} \right)^{\frac{1}{q}} \leqslant 2^{\frac{\Delta}{p}}
\end{align}
and
\begin{align}\label{eq:thm:metric_hardcore_2}
\left( \mathbb{E}_{x\leftarrow V}\left| \A(x)-\A'(x) \right|^{p}  \right)^{\frac{1}{p}} = O\left( \sqrt{\frac{p}{\ell}}\right) \text{ for } \A \text{ of complexity }\ell \text{ r. t. } \cA' 
\end{align}
The first follows by noticing that the quantity is convex with respect to $Y|E\in \cC$. Thus, the maximum is attained at one of extreme points which is, in this case, a flat distribution. The second fact follows from \Cref{lemma:lp_approximation}.

\section{Proof of \Cref{thm:metric_hardcore}}\label{app:thm:metric_hardcore}
To finish the proof it remains to justify the estimates
\begin{align}\label{eq:thm:metric_hardcore_1}
\left( \mathbb{E}_{x\leftarrow U}\left( 2^{n}\mathbf{P}_{Y|E}(x)\right)^{q} \right)^{\frac{1}{q}} \leqslant 2^{\frac{\Delta}{p}}
\end{align}
and
\begin{align}\label{eq:thm:metric_hardcore_2}
\left( \mathbb{E}_{x\leftarrow V}\left| \A(x)-\A'(x) \right|^{p}  \right)^{\frac{1}{p}} = O\left( \sqrt{\frac{p}{\ell}}\right) \text{ for } \A \text{ of complexity }\ell \text{ r. t. } \cA' 
\end{align}
The first follows by noticing that the quantity is convex with respect to $Y|E\in \cC$. Thus, the maximum is attained at one of extreme points which is, in this case, a flat distribution. The second fact follows from \Cref{lemma:lp_approximation}.

\section{Time-Success Ratio for Auxiliary Input Simulator Analysis of Stream Ciphers}\label{app:leakage-resilient}
\subsection{Preliminaries}
Weak pseudorandom functions, are \emph{ indistinguishable} from random functions, when queried on random inputs and fed with iniform secret key.
\begin{definition}[Weak pseudorandom functions]
A function $\F: \{0, 1\}^{k} \times \{0, 1\}^{n} \rightarrow \{0, 1\}^{m}$ is an $(\epsilon, s, q)$-secure weak PRF if its outputs on $q$ random inputs are indistinguishable from random by any distinguisher of size $s$, that is 
\begin{align*}
\left| \Pr \left[\D\left(\left( X_i \right)_{i=1}^{q},\F((K,X_i)_{i=1}^{q} \right)=1\right] - \Pr \left[\D\left(\left(X_i\right)_{i=1}^{q},\left(R_i\right)_{i=1}^{q}\right)=1 \right] \right| \leqslant \epsilon
\end{align*}
where the probability is over the choice of the random $X_i \leftarrow \{0,1\}^n$, the choice of a random key $K \leftarrow \{0,1\}^k$ and $R_i \leftarrow \{ 0,1\}^m$ conditioned on $R_i = R_j$ if $X_i = X_j$ for some $j < i$.
\end{definition}
Stream ciphers generate a keystream in a recursive manner. The security means that the output stream should be indistinguishable from uniform\footnote{We note that in a more standard notion the entire stream $X_1,\ldots,X_{q}$ is indistinguishable from random. This is implied by the notion above by a standard hybrid argument, with a loss of a multiplicative factor of $q$ in the distinguishing advantage.}.
\begin{definition}[Stream ciphers]
A \emph{stream-cipher} $\mathsf{SC} : \{0, 1\}^k \rightarrow \{0, 1\}^k \times \{0, 1\}^n$ is a function that need to be initialized with a secret state $S_0 \in \{0, 1\}^k$ and produces a sequence of output blocks $X_1, X_2, . . . $ computed as
\begin{align*}
 (S_i, X_i) := \mathsf{SC}(S_{i-1}).
\end{align*}
A stream cipher  $\mathsf{SC}$ is $(\epsilon,s,q)$-secure if for all $1 \leqslant i \leqslant q$, the random variable $X_i$ is $(s,\epsilon)$-pseudorandom given $X_1, . . . , X_{i-1}$ (the probability is also over the choice of the initial random key $S_0$).
\end{definition}
Now we define the security of leakage resilient stream ciphers, which follow the ``only computation leaks'' assumption.
\begin{definition}[Leakage-resilient stream ciphers]
A leakage-resilient stream-cipher is $(\epsilon,s,q,\lambda)$-secure if it is $(\epsilon,s,q)$-secure as defined above, but where the distinguisher in the $j$-th round gets $\lambda$ bits of arbitrary deceptively chosen leakage about the secret state accessed during this round. More precisely, before $(S_j,X_j) := \mathsf{SC}(S_{j−1})$ is computed, the distinguisher can choose any leakage function $f_j$ with range $\{0,1\}^{\lambda}$, and then not only get $X_j$, but also $\Lambda_j := f_j(\hat{S}_{j−1})$, where $\hat{S}_{j−1}$ denotes the part of the secret state that was modified (i.e., read and/or overwritten) in the computation $\mathsf{SC}(S_{j−1})$.
\end{definition}
Finally, we recall the standard notion of time-success ratio. It is very useful in quantifying how much security is transformed from the underlying primitive to the constructed object by the reduction.

\begin{definition}[Time-Success Ratio]\label{def:time-success}
We say that a cryptographic protocol has $k$ bits of security (or that it is $2^{k}$-secure) if for every $s$ and any adversary $\A$ of size $s$ the advantage $\A$ (probability of winning in the security game) is at most $\epsilon \leqslant s/2^{k}$.  
\end{definition}
\subsection{Time-Success Ratio Analysis}
Suppose that we have a simulator which guarantees if we have a simulator with complexity $t_{h} = O\left(t\cdot A\epsilon^{-\alpha}+B\epsilon^{-\beta}\right)$ then, according to \cite{JetchevP14}, we have a $(s',\epsilon',q)$-secure stream cipher where
\begin{align}\label{eq:transformation}
 \epsilon' = O\left( q\cdot \sqrt{2^{\lambda}\epsilon} \right),\quad s' =   \Omega\left( s\cdot A^{-1}(\epsilon')^{\alpha}\right)-A^{-1}B(\epsilon')^{\alpha-\beta}
\end{align}
Suppose that we want to prove $2^{k'}$-security in the sense of \Cref{def:time-success}. That is, we need to prove $s'/\epsilon' \geqslant 2^{k'}$ for every time-advantage pair $(s',\epsilon')$ such that $s'\geqslant 1$, where $k'$ is possibly big. Note that for a weak PRF we can assume the security $s\approx 2^{k}\epsilon$ for every $\epsilon$, that is that the best attack is by a brute-force search over the key space (see \cite{JetchevP14} for more justifiction). One can argue that, under the transformation \eqref{eq:transformation}, the worst-case adversary profile is when $\epsilon' \approx 2^{-k'}$ and $s'\approx 1$. Pugging this in \Cref{eq:transformation}, and using the fact that $s'\geqslant 1$ we obtain
\begin{align*}
2^{k}\cdot 2^{-2k'-\lambda} > A\cdot \left(2^{-k'}\right)^{-\alpha} + B\cdot \left(2^{-k'}\right)^{-\beta}.
\end{align*}
Substituting different values of $A,B,\alpha,\beta$  which correspond to the particular bounds, we get the values in \Cref{table:stream_ciphers_strength}.

\end{document}